\documentclass[10pt]{bmc_article}

\usepackage{cite} 
\usepackage{url}  
\usepackage{ifthen}  
\usepackage{multicol}   
\usepackage[utf8]{inputenc} 
\usepackage{amsmath}
\usepackage{bm}
\usepackage[pdftex]{graphicx}

\usepackage{xcolor}

\definecolor{orange}{rgb}{0.9,0.4,0}
\definecolor{maccha}{rgb}{0.1,0.4,0}

\newboolean{publ}

\newenvironment{bmcformat}{\begin{raggedright}\baselineskip20pt\sloppy\setboolean{publ}{false}}{\end{raggedright}\baselineskip20pt\sloppy}

\begin{document}
\begin{bmcformat}

\parindent = 20pt

\title{%
  Groupwise information sharing promotes ingroup favoritism in indirect reciprocity%
}

\author{%
  Mitsuhiro Nakamura$^1$%
  \and
  Naoki Masuda\correspondingauthor$^1$%
  \email{Naoki Masuda\correspondingauthor - masuda@mist.i.u-tokyo.ac.jp}%
}

\address{%
    \iid(1)Department of Mathematical Informatics,%
    The University of Tokyo,%
    7-3-1 Hongo, Bunkyo, Tokyo 113-8656, Japan%
}%

\maketitle


\begin{abstract}
\paragraph*{Background:}
Indirect reciprocity is a mechanism for cooperation in social dilemma
situations.
In indirect reciprocity, an individual is motivated to help another to acquire a
good reputation and receive help from others afterwards.
Another aspect of human cooperation is ingroup favoritism, whereby individuals
help members in their own group more often than those in other groups.
Ingroup favoritism is a puzzle for the theory of cooperation because it is not
easily evolutionarily stable.
In the context of indirect reciprocity, ingroup favoritism has been shown to be
a consequence of employing a double standard when assigning reputations to
ingroup and outgroup members.
An example of such a double standard is the situation in which helping an
ingroup member is regarded as good, whereas the same action toward an outgroup
member is regarded as bad.
\paragraph*{Results:}
We analyze a computational model of indirect reciprocity in which information
sharing is conducted groupwise.
In our model, individuals play social dilemma games within and across groups,
and the information about their reputations is shared within each group.
We show that evolutionarily stable ingroup favoritism emerges even if all the
players use the same reputation assignment rule regardless of group (i.e., a
single standard).
Two reputation assignment rules called simple standing and stern judging yield
ingroup favoritism; under these rules, cooperation with (defection against) good
individuals is regarded as good (bad) and defection against bad individuals is
regarded as good.
Stern judging induces much stronger ingroup favoritism than does simple
standing.
Simple standing and stern judging are evolutionarily stable against each other
when groups employing different assignment rules compete and the number of
groups is sufficiently large.
In addition, we analytically show as a limiting case that homogeneous
populations of reciprocators that use reputations are unstable when individuals
independently infer reputations of individuals, which is consistent with
previously reported numerical results.
\paragraph*{Conclusions:}
Our results suggest that ingroup favoritism can be promoted in indirect
reciprocity by the groupwise information sharing, in particular under the stern
judging assignment rule.
\end{abstract}

\ifthenelse{\boolean{publ}}{\begin{multicols}{2}}{}


\section*{Background}
\label{sec:introduction}


Behavioral nature of humans depends on the economy of reputations, where praise
and blame often lead to gain and loss of material
benefits~\cite{Sigmund2012,Pfeiffer2012}.
Humans, among other animals, cooperate via indirect reciprocity, which involves
cooperation beyond pairwise
relationships~\cite{Alexander1987,Trivers1971,Nowak1998a,Nowak1998b}.
In indirect reciprocity based on reputations, an individual acquires a good
reputation by behaving cooperatively in apposite situations.
The cost of maintaining a good reputation is compensated for by other
individuals' future cooperation toward the individual possessing the good
reputation.
Indirect reciprocity has been extensively studied in both theories~\cite{%
  Nowak1998a,Nowak1998b,%
  Leimar2001,Panchanathan2003,Fishman2003,Brandt2004,Ohtsuki2004,%
  Ohtsuki2006,Takahashi2006,Ohtsuki2007,%
  Chalub2006,Pacheco2006,%
  Ohtsuki2009,Uchida2010a,Uchida2010b%
} and experiments~\cite{%
  Wedekind2000,Semmann2004,Bolton2005,Sommerfeld2007,Pfeiffer2012}.


Another facet of human cooperation is that an individual often cooperates with
members in the same group and not with others, a phenomenon called ingroup
favoritism~\cite{%
  Tajfel1971,Yamagishi1998,Yamagishi1999,Kiyonari2000,%
  Goette2006,Yamagishi2008,Fowler2008,Rand2009,Guth2009,Mifune2010}.
Ingroup favoritism poses a puzzle for the theory of cooperation because it is
usually not Pareto efficient; i.e., the payoff to an individual in the case of
ingroup favoritism is smaller than that in the case of group-independent all-out
cooperation.
In addition, an individual implementing ingroup favoritism is worse off than an
individual defecting against both ingroup and outgroup members unless a specific
assumption is imposed.
In fact, known mechanisms for stable ingroup favoritism (e.g., correlation
between altruistic traits and phenotypic
tags~\cite{Jansen2006,Traulsen2008,Antal2009}, incomplete observability of
tags~\cite{Masuda2007}, combination of mutation of tags and limited
dispersal~\cite{Fu2012}) are, in our view, complicated.
Otherwise, stable ingroup favoritism requires an additional mechanism (e.g.,
intergroup conflict~\cite{Choi2007,Konrad2012}) that is capable of stabilizing
cooperation on its own.


If maintaining a good reputation is a concern, why do individuals want to
discriminate between ingroup and outgroup fellows?
One of the present authors has shown that ingroup favoritism is evolutionarily
stable in various situations when only group-level reputations are available in
regard to outgroup members~\cite{Masuda2012}.
In the model, an individual's action changes the individual's reputation in the
eyes of the ingroup members, and the action also changes the reputation of the
group to which the individual belongs.
It was revealed that the action rule of individuals (i.e., the strategy
depending on the reputation of the coplayer) toward ingroup and outgroup members
and the reputation assignment rule (also called the social norm) used for
evaluating ingroup and outgroup interactions, or at least the latter, must
discriminate between ingroup and outgroup members for stabilizing ingroup
favoritism.
An example is a rule whereby cooperation toward outgroup members is frowned
upon, whereas the same behavior toward ingroup members leads to a good
reputation.
Consistent with this theoretical example, Yamagishi and colleagues had conducted
behavioral experiments suggesting that ingroup favoritism occurs because
subjects anticipate that the reputation mechanism is functional only inside the
group~\cite{%
Yamagishi1998,Yamagishi1999,Kiyonari2000,Yamagishi2008,Mifune2010}.
These theoretical and experimental results suggest that double standards, in
terms of the action rule or the reputation assignment rule, may underpin ingroup
favoritism.


In the context of indirect reciprocity, group structure may play a crucial role
in spreading reputations of individuals via rumor and gossip.
In general, individuals interact more frequently with ingroup members than
with outgroup members~\cite{Fortunato2010}.
Therefore, rumor and gossip may enable sharing of reputations of individuals
more smoothly within a group than between different groups.
Most theoretical studies of indirect reciprocity have assumed that information
sharing and interactions occur randomly in a well-mixed population.
Otherwise, individuals are assumed to not exchange information about reputations
~\cite{Leimar2001,Brandt2004,Takahashi2006,Uchida2010b}.


In the present study, we explore a scenario of ingroup favoritism without
resorting to rules that apply double standards.
In practice, humans may not differentiate between ingroup and outgroup coplayers
with regard to their action rules or reputation assignment rules.
We analyze a group-structured model of indirect reciprocity, in which an
individual's reputation is shared by each group but not between groups.
We study the case in which all the players use the same reputation assignment
rule and the case in which players in different groups use different reputation
assignment rules.
We show that ingroup favoritism can emerge when players simply implement
reputation-based decision making and do not favor ingroup members.
Because of the assumed groupwise information sharing and some reputation
assignment error, ingroup and outgroup members tend to possess good and bad
reputations, respectively, without further assumptions.
In particular, ingroup favoritism is strong when individuals adopt a reputation
assignment rule called stern judging, under which helping bad individuals is
regarded as bad.


\section*{Methods}

\subsection*{Model}
\label{sec:model}


We consider an infinitely large population of players divided into $M$ ($\ge 2$)
groups.
Each group is assumed to contain the equal fraction, $1/M$, of players.
In the population, players are involved in sufficiently many rounds of the
so-called donation game.
In a one-shot donation game, two players are randomly selected from the
population, one as donor and the other as recipient.
We assume that the donor and recipient belong to the same group with probability
$\theta$.
The donor cooperates (C), i.e., provides help, or defects (D), i.e., refrains
from helping, depending on the donor's action rule and the recipient's
reputation (good (G) or bad (B)).
Action C imposes cost $c$ ($> 0$) on the donor and results in benefit $b$ ($>
c$) imparted to the recipient.
Action D does not change the payoff to either the donor or recipient.
A donor adopting action rule ALLC cooperates with any recipient.
A donor adopting action rule ALLD defects against any recipient.
A donor adopting action rule DISC cooperates with G recipients and defects
against B recipients.


To know a recipient's reputation, the donor consults the unique information
source, called the observer, that is shared by the group to which the donor
belongs.
Therefore, players in different groups may perceive different reputations (i.e.,
G or B) of the same player.
The observer in each group independently assigns a reputation to the donor and
shares it with the other players in the observer's group.
Observers intend the predefined reputation assignment toward a donor's action
but may assign a reputation opposite to the intention.
The $M$ observers independently commit such assignment error with probability
$\mu$ ($\ll 1$).
In the example of intragroup interaction shown in
Fig.~\ref{fig:example-observation}, all the three observers intended to assign G
to the donor, and one observer erroneously assigned B to the donor.
If the assignment error occurs, the ``wrong'' reputation is shared by all the
players in the group to which the observer belongs.


Observers assign reputations according to a common reputation assignment rule
unless otherwise stated.
We principally compare three rules: image scoring (IM), simple standing (ST),
and stern judging (JG)~\cite{%
  Nowak1998a,Nowak1998b,%
  Brandt2004,Ohtsuki2004,Ohtsuki2006,%
  Ohtsuki2007,Uchida2010a%
}, symbolically shown in Fig.~\ref{fig:assessment}.
Among the three rules, IM is the simplest rule under which observers assign G
and B to a donor that has selected C and D, respectively.
ST and JG are simplest among the so-called ``leading eight'' reputation
assignment rules that stabilize cooperation in well-mixed
populations~\cite{Ohtsuki2004,Ohtsuki2006}.
Under ST and JG, the new reputation of the donor depends on the action of the
donor (i.e., C or D) and the reputation of the recipient (i.e., G or B).
When a recipient has a G reputation, observers assign G and B to a C and D
donor, respectively, under both ST and JG.
When a recipient has a B reputation, observers assign G to a D donor under both
rules.
The two rules are different in that helping bad individuals (i.e., a donor's C
with a B recipient) is appreciated (i.e., G imparted by the observer) under ST,
whereas the same action of the donor is punished (i.e., B imparted by the
observer) under JG; JG is sterner than ST~\cite{Uchida2010a,Sigmund2012}.


After sufficiently many rounds of the donation game involving reputation
updates, the reputation distribution in the eyes of each group-specific observer
reaches a unique equilibrium.
In the equilibrium, we measure the quantities of interest such as the fractions
of G players, the probability of cooperation, and their dependence on groups.



\subsection*{Analysis Methods}
\label{sec:methods}

\subsubsection*{Equilibria of the reputation dynamics}
\label{sec:methods:equilibria}

Table~\ref{tbl:definitions} summarizes the definitions of the symbols used in
this section.


We examine the stability of a homogeneous population of DISC players.
Each player bears a reputation vector, ${\bm r} = (r_1, r_2, \dots, r_M) \in
\{\rm G, B\}^M$, in the eyes of $M$ observers, each representing a group.
We denote by $p_k({\bm r})$ the probability that a player in group $k$ has
reputation vector $\bm r$.
By adopting the formalism developed by Ohtsuki \& Iwasa~\cite{Ohtsuki2004}, we
obtain the following reputation dynamics:
\begin{equation}
  \label{eq:basic-reputation-dynamics}
  \frac{{\rm d}}{{\rm d}t}p_k({\bm r}) =
  - p_k({\bm r})
  + \sum_{{\bm r}^\prime \in \{\rm G, B\}^M}
  \left\{ \theta p_k({\bm r}^\prime) +
    (1-\theta) p_{-k}({\bm r}^\prime) \right\}
  \prod_{k^\prime=1}^M \Phi_{r_{k^\prime}} (\sigma(r^\prime_k), r^\prime_{k^\prime}).
\end{equation}
The summation on the right-hand side of Eq.~(\ref{eq:basic-reputation-dynamics})
represents the average over the recipient's reputation vector ${\bm r}^\prime$.
With probability $\theta$, a game involves a donor and a recipient in group $k$,
and the recipient has reputation vector ${\bm r}^\prime$ with probability
$p_k({\bm r}^\prime)$.
With probability $1-\theta$, a donor and a recipient belong to group $k$ and
another group, respectively, and the recipient has reputation vector ${\bm
  r}^\prime$ with probability $p_{-k}({\bm r}^\prime) \equiv \sum_{k^\prime = 1,
  k^\prime \neq k}^M p_{k^\prime}({\bm r}^\prime) / (M-1)$.
$\sigma(r^\prime_k)$ represents a donor's action toward a recipient having
reputation $r^\prime_k$.
Because we assume DISC donors, $\sigma ({\rm G}) = {\rm C}$ and $\sigma ({\rm
  B}) = {\rm D}$.
The reputation assignment rule is essentially given by
$\Phi_{r_{k^\prime}}(\sigma(r^\prime_k), r^\prime_{k^\prime}) \in \{1-\mu,
\mu\}$, which is the probability that an observer in group $k^\prime$ assigns
reputation $r_{k^\prime}$ to a donor in group $k$.
This probability depends on the donor's action $\sigma(r^\prime_k)$ toward a
recipient having reputation $r^\prime_{k^\prime}$ in the eyes of the observer in
each group $k^\prime$.
In Tab.~\ref{tbl:assessment}, we list the $\Phi$ values under different
assignment rules.
It should be noted that all the observers use a unique assignment rule unless
otherwise stated; we do not basically assume that observers employ different
assignment rules as in previous
studies~\cite{Brandt2004,Chalub2006,Pacheco2006,Uchida2010a,Uchida2010b}.


We reduce Eq.~(\ref{eq:basic-reputation-dynamics}) to mean field dynamics of two
reputation distributions.
First, we apply summation $\sum_{{\bm r}_{-k}} \equiv
\sum_{r_1}\sum_{r_2}\dots\sum_{r_{k-1}}\sum_{r_{k+1}}\dots \sum_{r_M}$ to both
sides of Eq.~(\ref{eq:basic-reputation-dynamics}) to obtain the reputation
dynamics in the eyes of ingroup observers as follows:
\begin{equation}
  \label{eq:ingroup-reputation-dynamics}
  \frac{{\rm d}}{{\rm d}t}p_{\rm in}(r) =
  - p_{\rm in}(r)
  + \sum_{r^\prime \in \{\rm G, B\}}
  \left\{ \theta p_{\rm in}(r^\prime) +
    (1-\theta) p_{\rm out}(r^\prime) \right\}
  \Phi_r (\sigma(r^\prime), r^\prime),
\end{equation}
where
$p_{\rm in}(r) \equiv \sum_{{\bm r}_{-k}} p_k({\bm r})$ and
$p_{\rm out}(r) \equiv \sum_{{\bm r}_{-k}} p_{-k}({\bm r})$
are the probabilities that a player has reputation $r \in \{\rm G, B\}$ in the
eyes of ingroup and outgroup observers, respectively.
The two terms inside the curly brackets on the right-hand side of
Eq.~(\ref{eq:ingroup-reputation-dynamics}) correspond to the two situations
shown in Fig.~\ref{fig:example-interactions}{\bf (a)} and {\bf (b)}.
With probability $\theta p_{\rm in}(r^\prime)$, the recipient belongs to the
donor and observer's group, and has reputation $r^\prime$
(Fig.~\ref{fig:example-interactions}{\bf (a)}).
With probability $(1-\theta) p_{\rm out}(r^\prime)$, the recipient does not
belong to the donor and observer's group, and has reputation $r^\prime$
(Fig.~\ref{fig:example-interactions}{\bf (b)}).


Second, by applying summation $\sum_{{\bm r}_{-\ell}} \equiv
\sum_{r_1}\sum_{r_2}\dots\sum_{r_{\ell-1}}\sum_{r_{\ell+1}}\dots \sum_{r_M}$,
where $\ell \neq k$, to both sides of Eq.~(\ref{eq:basic-reputation-dynamics}),
we obtain
\begin{equation}
  \label{eq:outgroup-reputation-dynamics-before}
  \begin{split}
    \frac{\rm d}{{\rm d}t}p_k(r_\ell) &=
    - p_k(r_\ell) \;+ \sum_{r^\prime_k \in \{\rm G, B\}}\sum_{r^\prime_\ell \in \{\rm G, B\}} \\
    & \left\{ \theta p_k(r^\prime_k, r^\prime_\ell) +
      (1-\theta) \left[
        \frac{1}{M-1}\; p_\ell(r^\prime_k, r^\prime_\ell) +
        \left(1-\frac{1}{M-1}\right)\; p_{-k\ell}(r^\prime_k, r^\prime_\ell)
      \right]\right\}
    \Phi_{r_\ell} (\sigma(r^\prime_k), r^\prime_\ell),
  \end{split}
\end{equation}
where
$p_k(r^\prime_k, r^\prime_\ell)$,
$p_\ell(r^\prime_k, r^\prime_\ell)$, and
$p_{-k\ell}(r^\prime_k, r^\prime_\ell)$
are the probabilities that a player in group $k$, group $\ell$, and a group
other than $k$ and $\ell$, respectively, has reputation $r^\prime_k$ and
$r^\prime_\ell$ in the eyes of observers in groups $k$ and $\ell$.
By approximating the three probabilities by
$p_{\rm in}(r^\prime_k)p_{\rm out}(r^\prime_\ell)$,
$p_{\rm out}(r^\prime_k)p_{\rm in}(r^\prime_\ell)$, and
$p_{\rm out}(r^\prime_k)p_{\rm out}(r^\prime_\ell)$,
respectively, we obtain the mean-field reputation dynamics in the eyes of
outgroup observers as follows:
\begin{equation}
  \label{eq:outgroup-reputation-dynamics}
  \begin{split}
    \frac{\rm d}{{\rm d}t}p_{\rm out}(r) &=
    - p_{\rm out}(r) \;+ \sum_{r^\prime \in \{\rm G, B\}}
    \sum_{r^{\prime\prime} \in \{\rm G, B\}} \\
    & \left\{ \theta p_{\rm in}(r^\prime)p_{\rm out}(r^{\prime\prime}) +
      (1-\theta) \left[
        \frac{1}{M-1}\; p_{\rm out}(r^\prime)p_{\rm in}(r^{\prime\prime}) +
        \left(1-\frac{1}{M-1}\right)\; p_{\rm out}(r^\prime)p_{\rm out}(r^{\prime\prime})
      \right]\right\} \times \\
    & \Phi_r (\sigma(r^\prime), r^{\prime\prime}).
  \end{split}
\end{equation}
The three terms inside the curly brackets on the right-hand side of
Eq.~(\ref{eq:outgroup-reputation-dynamics}) correspond to the three situations
shown in Fig.~\ref{fig:example-interactions}{\bf (c)}, {\bf (d)}, and {\bf
  (e)}.
With probability $\theta p_{\rm in}(r^\prime) p_{\rm out}(r^{\prime\prime})$,
the recipient belongs to the donor's group, which differs from the observer's
group, and has reputation $r^\prime$ and $r^{\prime\prime}$ in the eyes of the
donor and observer, respectively (Fig.~\ref{fig:example-interactions}{\bf
  (c)}).
With probability $(1-\theta)[1/(M-1)] p_{\rm out}(r^\prime) p_{\rm
  in}(r^{\prime\prime})$, the recipient belongs to the observer's group, which
differs from the donor's group, and has reputation $r^\prime$ and
$r^{\prime\prime}$ in the eyes of the donor and observer, respectively
(Fig.~\ref{fig:example-interactions}{\bf (d)}).
With probability $(1-\theta)[1-1/(M-1)] p_{\rm out}(r^\prime) p_{\rm
  out}(r^{\prime\prime})$, the recipient belongs to a group different from the
donor's and observer's groups, and has reputation $r^\prime$ and
$r^{\prime\prime}$ in the eyes of the donor and observer, respectively
(Fig.~\ref{fig:example-interactions}{\bf (e)}).


By setting ${\rm d} p_{\rm in}(r) / {\rm d}t = {\rm d} p_{\rm out}(r) / {\rm d}t
= 0$ in Eqs.~(\ref{eq:ingroup-reputation-dynamics}) and
(\ref{eq:outgroup-reputation-dynamics}), we identify stationary points that are
candidates of stable equilibria of the reputation dynamics.
We examine the conditions $\det {\bm J} > 0$ and ${\rm Tr}\; {\bm J} < 0$, where
$\bm J$ is the Jacobian matrix, at each stationary point to identify all the
stable equilibria.
We confirmed that the stable equilibrium denoted by $p^*_{\rm in}(r)$ and
$p^*_{\rm out}(r)$ is unique under each reputation assignment rule.


\subsubsection*{Stability against invasion by ALLC and ALLD mutants}
\label{sec:methods:stability}


We check the evolutionary stability of a homogeneous population composed of DISC
players against invasion by an infinitesimal fraction of mutants adopting ALLC
or ALLD.
The payoff to a DISC resident player is given by
\begin{equation}
  \label{eq:payoff-DISC}
  \pi_{\rm DISC} = (b-c) \left[\theta p_{\rm in}^*({\rm G}) +
    (1-\theta) p_{\rm out}^*({\rm G})\right],
\end{equation}
and those to ALLC and ALLD mutants are given by
\begin{equation}
  \label{eq:payoff-ALLC}
  \pi_{\rm ALLC} = -c + b \left[\theta p_{\rm in}^{\rm C}({\rm G}) +
    (1-\theta) p_{\rm out}^{\rm C}({\rm G}) \right]
\end{equation}
and
\begin{equation}
  \label{eq:payoff-ALLD}
  \pi_{\rm ALLD} = b \left[\theta p_{\rm in}^{\rm D}({\rm G}) +
    (1-\theta) p_{\rm out}^{\rm D}({\rm G}) \right],
\end{equation}
respectively.
In Eqs.~(\ref{eq:payoff-ALLC}) and (\ref{eq:payoff-ALLD}), $p_{\rm in}^{\rm
  C}({\rm G})$, $p_{\rm out}^{\rm C}({\rm G})$, $p_{\rm in}^{\rm D}({\rm G})$, and
$p_{\rm out}^{\rm D}({\rm G})$ represent the probabilities that the mutants
selecting C and D acquire G reputations in the eyes of ingroup and outgroup
observers, and are given by
\begin{equation}
  \nonumber
  p_{\rm in}^a({\rm G}) = \sum_{r^\prime \in \{\rm G, B\}}
  \left\{ \theta p_{\rm in}^*(r^\prime) +
    (1-\theta) p_{\rm out}^*(r^\prime) \right\}
  \Phi_{\rm G} (a, r^\prime)
\end{equation}
and
\begin{equation}
  \nonumber
  \begin{split}
    p_{\rm out}^a({\rm G}) &=
    \sum_{r^\prime \in \{\rm G, B\}} \sum_{r^{\prime\prime} \in \{\rm G, B\}} \\
    &\left\{ \theta p_{\rm in}^*(r^\prime)p_{\rm out}^*(r^{\prime\prime}) +
      (1-\theta)\left[ \frac{1}{M-1} p_{\rm out}^*(r^\prime)p_{\rm
          in}^*(r^{\prime\prime}) + (1-\frac{1}{M-1}) p_{\rm
          out}^*(r^\prime)p_{\rm out}^*(r^{\prime\prime})
      \right]\right\} \times \\
   & \Phi_{\rm G} (a, r^{\prime\prime}),
  \end{split}
\end{equation}
where $a = {\rm C}$ or ${\rm D}$.
The population of DISC players is stable against invasion by ALLC and ALLD
mutants if
\begin{equation}
  \label{eq:stability-condition}
  \pi_{\rm DISC} > \max \left\{ \pi_{\rm ALLC}, \pi_{\rm ALLD} \right\}.
\end{equation}


\subsubsection*{Cooperativeness}
\label{sec:methods:cooperativeness}


DISC donors cooperate exclusively with G recipients. Therefore, in each stable
equilibrium, the probability of cooperation, which we call the cooperativeness,
toward ingroup and outgroup recipients is given by $p^*_{\rm in}({\rm G})$ and
$p^*_{\rm out}({\rm G})$, respectively.
The cooperativeness for the entire population is given by
\begin{equation}
  \label{eq:cooperativeness}
  \psi \equiv \theta p^*_{\rm in}({\rm G}) +
  (1-\theta) p^*_{\rm out}({\rm G}).
\end{equation}


\subsubsection*{Measurement of ingroup bias}
\label{sec:methods:ingroup-bias}


To quantify the degree of ingroup bias, we measure the difference between
ingroup and outgroup cooperativeness, defined by
\begin{equation}
  \label{eq:ingroup-bias}
  \rho \equiv p^*_{\rm in}({\rm G}) - p^*_{\rm out}({\rm G}).
\end{equation}
When $\rho \approx -1$, players basically cooperate with outgroup recipients and
defect against ingroup recipients, implying outgroup favoritism.
When $\rho \approx 0$, players equally likely cooperate with ingroup and
outgroup recipients.
When $\rho \approx 1$, players cooperate with ingroup recipients and defect
against outgroup recipients, implying ingroup favoritism.



\section*{Results}
\label{sec:results}


Table~\ref{tbl:equilibria} summarizes the results obtained under the three
reputation assignment rules.
It shows the stable fractions of G players in the eyes of ingroup and outgroup
observers (i.e., $p^*_{\rm in}(\rm G)$ and $p^*_{\rm out}(\rm G)$), the
stability conditions, cooperativeness (i.e., $\psi$), and the degree of ingroup
bias (i.e., $\rho$).



\subsection*{IM}
\label{sec:results:IM}


Under IM, the equilibrium fractions of G players in the eyes of ingroup and
outgroup observers are both equal to $\psi = 1/2$.
Therefore, ingroup favoritism does not occur, i.e., $\rho = 0$.
Furthermore, the population of DISC players is invaded by ALLC mutants such that
it is unstable.
This result is consistent with the established result that cooperation is
usually unstable under IM because observers do not distinguish between selfish
defection (i.e., D against G recipients) and justified defection (i.e., D
against B recipients)~\cite{Panchanathan2003,Brandt2004,Nowak2005,Sigmund2012}.



\subsection*{ST}
\label{sec:results:ST}


Under ST, DISC players almost always cooperate with ingroup recipients, i.e.,
$p^*_{\rm in}({\rm G}) = 1-\mu$.
This result is consistent with the previous results in which ST enables perfect
cooperation when a population does not possess group structure (corresponding to
$M = 1$)~\cite{Ohtsuki2004,Ohtsuki2006}.


The fraction of G players in the eyes of outgroup observers is given by
\begin{equation}
  p^*_{\rm out}({\rm G}) = 1 - \mu\frac{1+\theta}{\theta} + O(\mu^2).
\end{equation}
Therefore, DISC players almost always cooperate with both ingroup and outgroup
recipients unless $\theta$ is small (i.e., $\psi = 1 - \mu/\theta + O(\mu^2)$).
Because donors defect slightly more often against outgroup than ingroup
recipients, weak ingroup favoritism occurs (i.e., $\rho = \mu/\theta +
O(\mu^2)$).


Equations~(\ref{eq:payoff-DISC}), (\ref{eq:payoff-ALLC}), and
(\ref{eq:payoff-ALLD}) yield the payoff differences given by
\begin{equation}
  \label{eq:relative-payoff-ALLC-ST}
  \pi_{\rm ALLC} - \pi_{\rm DISC} = \frac{\mu}{\theta}
  \left[ b(1-\theta) - c \right] + O(\mu^2)
\end{equation}
and
\begin{equation}
  \label{eq:relative-payoff-ALLD-ST}
  \pi_{\rm ALLD} - \pi_{\rm DISC} = -(b-c) + O(\mu).
\end{equation}
Therefore, the stability condition (Eq.~(\ref{eq:stability-condition})) reads
\begin{equation}
  \label{eq:stability-condition-ST}
  1 < \frac{b}{c} < \frac{1}{1-\theta}.
\end{equation}
ALLC mutants invade DISC players if $b/c > 1/(1-\theta)$.
The cooperation is stable up to a large value of $b/c$ when ingroup interaction
is frequent (i.e., large $\theta$).
ALLD mutants invade a DISC population under a trivial condition $b/c < 1$.



\subsection*{JG}
\label{sec:results:JG}


Under JG, DISC players have the same cooperativeness toward ingroup recipients
as under ST, i.e., $p^*_{\rm in}({\rm G}) = 1-\mu$.
This result is consistent with the previous results in which JG enables perfect
cooperation when a population does not possess group structure (corresponding to
$M = 1$)~\cite{Ohtsuki2004,Ohtsuki2006}.


The fraction of G players in the eyes of outgroup observers is given by
\begin{equation}
  p^*_{\rm out}({\rm G}) = \frac{1}{2}.
\end{equation}
Therefore, DISC players cooperate with outgroup recipients with probability
$1/2$.
In contrast to the case of ST, frequent intergroup interaction considerably
reduces cooperation under JG (i.e., $\psi = (1+\theta)/2 + \mu\theta$).
The degree of ingroup bias under JG is given by $\rho = 1/2 - \mu$, which is
independent of $\theta$.
DISC players show a significant level of ingroup favoritism, even though they
simply use the reputations without intending to discriminate recipients by the
group identity.


The payoff differences are given by
\begin{equation}
  \label{eq:relative-payoff-ALLC-JG}
    \pi_{\rm ALLC} - \pi_{\rm DISC} =
    -\frac{1-\theta}{2} \left[ b\frac{M\theta-1}{M-1} + c \right] + O(\mu)
  \end{equation}
and
\begin{equation}
  \label{eq:relative-payoff-ALLD-JG}
  \pi_{\rm ALLD} - \pi_{\rm DISC} =
  -\frac{1}{2}\left[
    b\frac{1+(M-3)\theta+M\theta^2}{M-1} -c(1+\theta) \right] + O(\mu).
\end{equation}
The stability condition reads
\begin{equation}
  \label{eq:stability-condition-JG}
  \begin{cases}
    \displaystyle
    \frac{(M-1)(1+\theta)}{1+(M-3)\theta+M\theta^2} < \frac{b}{c}
    < \frac{M-1}{1-M\theta} &
    \displaystyle
    \text{if } 0 \le \theta < \frac{1}{M}, \\
    \displaystyle
    \frac{(M-1)(1+\theta)}{1+(M-3)\theta+M\theta^2} < \frac{b}{c} &
    \displaystyle
    \text{if } \frac{1}{M} \le \theta \le 1.
  \end{cases}
\end{equation}
The DISC population is resistant to invasion by ALLC mutants when $\theta \ge
1/M$, i.e., when ingroup interaction occurs more frequently than in the case of
unbiased random pairing.
When $\theta < 1/M$ and $b/c > (M-1)/(1-M\theta)$, ALLC mutants invade the
population of DISC players.
When $b/c < (M-1)(1+\theta)/[1+(M-3)\theta+M\theta^2]$, ALLD mutants invade the
population of DISC players.
The cooperation is stable down to a small value of $b/c$ if ingroup interaction
is frequent (i.e., large $\theta$) or the number of groups (i.e., $M$) is small.
In the limit $M \to \infty$, Eq.~(\ref{eq:stability-condition-JG}) is reduced to
$b/c > 1/\theta$, which coincides with the results obtained from a previous
model with infinite groups~\cite{Masuda2012}.


Under both ST and JG, in particular JG, ingroup favoritism emerges.
This is because the donors (equivalently, ingroup observers) and outgroup
observers generally perceive different reputations of the same player due to the
assignment error (see Figs.~\ref{fig:example-observation},
\ref{fig:example-interactions}{\bf(c)}, \ref{fig:example-interactions}{\bf(d)}, and \ref{fig:example-interactions}{\bf(e)}).
For example, if a donor defects against a recipient whose reputation is B in the
eyes of the donor's group members, the donor receives a G reputation from the
ingroup observer.
However, if the same recipient has a G reputation in the eyes of the outgroup
observer, the outgroup observer assigns B to the donor under ST and JG.
As another example, if a donor cooperates with a recipient whose reputation is G
in the eyes of the donor's group members, the donor receives G from the ingroup
observer.
However, if the recipient has a B reputation in the eyes of the outgroup
observer, the outgroup observer assigns B to the donor under JG.
As these examples suggest, different groups may perceive the opposite
reputations of the same players in a long run.
Players in the same group coordinate the subjective information about a given
player's reputation, whereas those in different groups do not.
This discrepancy causes ingroup favoritism.


\subsection*{Numerical Results}
\label{sec:results:numerical}


%
We compare the theoretical results with numerical results obtained from
individual-based simulations in Fig.~\ref{fig:equilibria}.
The procedure of the numerical analysis is described in
Appendix~\ref{appendix:numerical-homo}.
%
The analytical and numerical results are sufficiently close to each other in
terms of both cooperativeness (Fig.~\ref{fig:equilibria}{\bf (a)}) and ingroup
bias (Fig.~\ref{fig:equilibria}{\bf (b)}).


%
We also examine the error-prone case in which donors fail to help recipients
(i.e., select D when the donors intend C) with probability
$\epsilon$~\cite{Fishman2003}.
%
%
The numerical results for $\epsilon = 0.01$ and $0.1$ are shown in
Fig.~\ref{fig:equilibria-eps}.
%
%
The error reduces cooperativeness (Fig.~\ref{fig:equilibria-eps}{\bf (a)}) and
ingroup bias (Fig.~\ref{fig:equilibria-eps}{\bf (b)}) under both ST and JG (see
Fig.~\ref{fig:equilibria} for the error-free case).
Nevertheless, the results with the error are qualitatively the same as those
without the error.



\subsection*{Mixed Assignment Rules}
\label{sec:results:mixed}


We have shown that JG leads to strong ingroup favoritism, whereas ST does not.
To examine the transition between the two regimes, we consider an assignment
rule denoted by MX, which is a mixture of JG and ST.
In a one-shot game under MX, observers independently assign reputations by using
JG with probability $\alpha$ and ST with probability $1-\alpha$.
Therefore,
$\Phi_{\rm G}({\rm C, G}) = 1-\mu$,
$\Phi_{\rm G}({\rm D, G}) = \mu$,
$\Phi_{\rm G}({\rm C, B}) = \alpha\mu+(1-\alpha)(1-\mu)$, and
$\Phi_{\rm G}({\rm D, B}) = 1-\mu$.
Parameter $\alpha$ controls the degree of sternness with which observers assign
B to donors that cooperate with B recipients.
ST and JG correspond to $\alpha = 0$ and $\alpha = 1$, respectively.
We numerically solve Eqs.~(\ref{eq:ingroup-reputation-dynamics}) and
(\ref{eq:outgroup-reputation-dynamics}) under MX.


The results under MX are shown in Fig.~\ref{fig:mixed}.
Sternness gradually decreases cooperativeness (Fig.~\ref{fig:mixed}{\bf (a)})
and increases ingroup bias (Fig.~\ref{fig:mixed}{\bf (b)}) for different values
of $M$ and $\theta$.
The results interpolate those for ST and JG and imply that sternness promotes
ingroup favoritism.
The shaded parameter regions in Fig.~\ref{fig:mixed}{\bf (c)}\---{\bf (f)}
indicate the values of $\alpha$ and $b/c$ for which DISC residents are stable.
Above (below) the shaded regions, ALLC (ALLD) mutants invade the DISC
population.
In all the cases, the upper and lower bounds of the stability region in terms of
$b/c$ increase with $\alpha$.
A decrease in $M$ induces cooperativeness and reduces ingroup bias.
A decrease in $M$ also broadens the stability regions if $\theta$ is large.
An increase in $\theta$ induces cooperativeness, reduces ingroup bias, and
broadens the stability regions for the following reason.
When $\theta$ is large, players are largely involved in ingroup interactions.
Then, they do not suffer from a B reputation that outgroup observers may
frequently attach to the donor because of the discrepancy between players'
reputations perceived by different groups (see subsection JG in Results for
related discussion).


\subsection*{Heterogeneous Assignment Rules}
\label{sec:results:heterogeneous}


We have assumed that all the groups use a common reputation assignment rule.
In this section, we numerically examine a case in which observers in different
groups use different reputation assignment rules.
We consider a situation in which $m$ ($1 \le m \le M-1$) groups use JG and $M-m$
groups use ST.
The procedure of the numerical analysis is described in
Appendix~\ref{appendix:numerical-hetero}.


Numerically obtained equilibria with $M = 8$ and $M = 20$ are shown in
Fig.~\ref{fig:equilibria-STJG}{\bf (a)} and {\bf (b)}, respectively.
As the number of JG groups (i.e., $m$) increases, the cooperativeness
($\psi_{\rm ST}$ and $\psi_{\rm JG}$ for ST and JG groups, respectively)
decreases, and ingroup bias ($\rho_{\rm ST}$ and $\rho_{\rm JG}$ for ST and JG
groups, respectively) increases.
Figure~\ref{fig:equilibria-STJG}{\bf (c)} and {\bf (d)} shows the difference
between the payoff to a player in a ST group and that to a player in a JG group
(i.e., $\pi_{\rm JG} - \pi_{\rm ST}$) when $M = 8$ and $M = 20$, respectively.
When the benefit-to-cost ratio is small (i.e., $b = 2$), $\pi_{\rm JG} -
\pi_{\rm ST}$ is positive.
Therefore, if observers update their assignment rules according to an
evolutionary dynamics (e.g., group competition~\cite{Pacheco2006}), the
evolutionary dynamics would lead to a homogeneous population in which all the
observers adopt JG.
When the benefit-to-cost ratio is large (i.e., $b = 6$), $\pi_{\rm JG} -
\pi_{\rm ST}$ is positive when $m$ is large and negative when $m$ is small.
This implies that a homogeneous population of ST and that of JG are bistable
under evolutionary dynamics.
The basin of attraction for the homogeneous ST population in terms of $m$
broadens as $b$ increases.
When the benefit-to-cost ratio takes an intermediate value (i.e., $b = 4$), the
results for $M=8$ (Fig.~\ref{fig:equilibria-STJG}{\bf (c)}) and those for
$M=20$ (Fig.~\ref{fig:equilibria-STJG}{\bf (d)}) are qualitatively different.
For $M=8$, $\pi_{\rm JG} - \pi_{\rm ST}$ is negative only when $m = 2$ or $3$.
Therefore, a stable mixture of ST and JG groups and a homogeneous population of
JG are bistable.
For $M=20$, a homogeneous population of ST and that of JG are bistable.



\section*{Discussion}
\label{sec:discussion}


In the present study, we showed that ingroup favoritism emerges in a
group-structured model of indirect reciprocity.
In our model, players share information about reputations in each group but not
across different groups.
We assumed that a player's action purely depends on the coplayer's reputation;
players do not refer to the group identity of the coplayers or use other types
of prejudice.
We also assumed that observers impartially assess ingroup and outgroup donors.
We analyzed the model using a mean-field approximation and numerical
simulations.
Ingroup favoritism occurs under both simple standing (ST) and stern judging (JG)
assignment rules.
The cooperativeness is reduced by the frequent intergroup interactions, i.e.,
small $\theta$.
The ingroup bias is severer and the cooperativeness is smaller under JG than
under ST.
The parameter region for the stability of the cooperative equilibrium is larger
under JG than under ST.
Under ST and JG, a population of discriminators is evolutionarily stable if the
probability of ingroup interaction (i.e., $\theta$) is sufficiently large.
If $\theta$ is small, the population is invaded by unconditional cooperators and
unconditional defectors under ST and JG, respectively.
We also studied the case in which observers may adopt different assignment rules
in different groups.
We found that JG would dominate ST in evolutionary settings when the
benefit-to-cost ratio is small.
Otherwise, the homogeneous population in which all the groups employ ST and that
in which all the groups employ JG are bistable in large parameter regions.


Different mechanisms govern ingroup favoritism in our model and that observed in
psychological experiments~\cite{%
  Tajfel1971,Yamagishi1998,Yamagishi1999,Kiyonari2000,%
  Goette2006,Yamagishi2008,Fowler2008,Rand2009,Guth2009,Mifune2010}.
In the latter, players use a cue that indicates the group identity of the
coplayer and preferably cooperate with ingroup members.
In our model, players do not refer to the group identity of the coplayer.
They show ingroup favoritism because they perceive that outgroup members have
bad reputations more often than do ingroup members.
%


We implemented the group structure by controlling probabilities of ingroup and
outgroup interactions (i.e., $\theta$ and $1-\theta$, respectively) and assuming
the groupwise information sharing.
In terms of the structure of information sharing, most previous theoretical
studies of indirect reciprocity are classified into two types: public~\cite{%
  Nowak1998a,Nowak1998b,Panchanathan2003,Fishman2003,Ohtsuki2004,Ohtsuki2006,%
  Chalub2006,Pacheco2006,Ohtsuki2007,Ohtsuki2009,Masuda2012%
} and private~\cite{%
  Leimar2001,Brandt2004,Takahashi2006,Ohtsuki2009,Uchida2010b%
} reputation models.


In public reputation models, all the players have access to a common information
source that provides the reputation values of the players.
Therefore, a donor and observer perceive the same reputation of a recipient such
that they do not suffer from the discrepancy of reputations.
In public reputation models without group structure of the population, ST and JG
realize evolutionarily stable cooperation~\cite{Ohtsuki2004,Ohtsuki2006}.
This result is consistent with ours because, in the limit $\theta \to 1$,
Eqs.~(\ref{eq:stability-condition-ST}) and (\ref{eq:stability-condition-JG}) are
reduced to a trivial condition $b/c > 1$ such that the population of
discriminators is stable under ST and JG.


In private reputation models, each player individually collects others'
reputations such that a reputation of a player varies between individuals.
In contrast to the case of public reputation models, a homogeneous population of
discriminators is invaded by unconditional cooperators in private reputation
models.
A mixture of discriminators and unconditional cooperators is often stable under
variants of ST~\cite{Leimar2001,Brandt2004,Takahashi2006,Uchida2010b}.
Under variants of JG, a population of discriminators is invaded by unconditional
defectors~\cite{Ohtsuki2009,Uchida2010b} (but see Ref.~\cite{Takahashi2006}), or
discriminators and unconditional cooperators are frequent in an island
model if dispersal of offspring is confined within each island~\cite{Brandt2004}.
In the limit $\theta \to 0$ and $M \to \infty$, our model can be interpreted as
a private reputation model.
In this situation, the population of discriminators is unstable because
Eqs.~(\ref{eq:stability-condition-ST}) and (\ref{eq:stability-condition-JG}) are
violated.
Therefore, the results obtained from our model in this limit are consistent with
the previous results.


For intermediate $\theta$ and $M$ values, our model uses a public reputation
scheme within each group and a private reputation scheme across groups.
In this sense, the structure of information sharing in our model is situated
between public and private reputation models.


One of the present authors previously studied a model of ingroup favoritism on
the basis of indirect reciprocity~\cite{Masuda2012}, which we refer to as the
multiple standard model.
The multiple standard model and the model analyzed in the present study are
different in two aspects.
First, in the multiple standard model, a given player's reputation is made
public to different groups such that the problem of coordination in regard to
reputations among different groups does not exist.
In the present model, observers in different groups may differently perceive a
player's reputation, which leads to the coordination problem.
Second, in the multiple standard model, observers are allowed to use different
rules to assign reputations to ingroup and outgroup members.
Similarly, donors may use different action selection rules toward ingroup and
outgroup recipients.
Then, ingroup favoritism of different degrees emerges.
Consider a situation in which the action rule is of a single standard such that
donors are discriminators toward both ingroup and outgroup recipients.
Then, at most partial ingroup favoritism in which players always cooperate with
ingroup members and partially (i.e., with probability $1/2$) cooperate with
outgroup members is evolutionarily stable.
Consider another situation in which the action rule is of a double standard such
that donors are discriminators toward ingroup members and unconditional
defectors toward outgroup members.
Then, perfect ingroup favoritism in which players always cooperate with ingroup
members and always defect against outgroup members is evolutionarily stable.
In the present model, observers use a single-standard reputation assignment
rule, and donors use a single-standard action rule.
Then, partial ingroup favoritism, but not perfect ingroup favoritism, can be
evolutionarily stable.


Group competition models of indirect reciprocity were previously
studied~\cite{Chalub2006,Pacheco2006}.
In references~\cite{Chalub2006,Pacheco2006}, the authors numerically examined
competition between different assignment rules employed in different groups.
In our terminology, they assumed that the donation game is played inside each
group and that reputations are updated exclusively by ingroup observers under
the public reputation scheme.
They showed that JG (stern-judging in their terminology) emerges in the course
of evolutionary dynamics based on group competition and individual selection.
Their models and ours are fundamentally different although both studies have
stressed the importance of JG.
First, they assumed group competition and we did not.
Second, they mainly focused on competition between different assignment rules
and we did not; we only studied the special case in which observers in different
groups adopt either of ST or JG.
Third, we determined the possibility of ingroup favoritism and group-independent
cooperation.
In contrast, their model is not concerned with ingroup favoritism because
interaction between a donor and recipient in different groups is not assumed.


Uchida and Sigmund analyzed competition between assignment rules by using
replicator dynamics~\cite{Uchida2010a}.
In their model, a player selected as donor uses the public information source
corresponding to the assignment rule that the player adopts.
For example, if the surviving assignment rules are only ST and JG (SUGDEN and
KANDORI, respectively, in their terminology), there are two public information
sources.
Although their model is apparently a public reputation model, the players can be
interpreted to belong to one of the groups defined by the assignment rule;
members in each group share a common information source and use the same
assignment rule.
Helping a recipient having a bad reputation in the eyes of both ST and JG groups
is assessed to be good by the ST group and bad by the JG group.
Therefore, JG players assess ST players to be bad more often than they assess JG
players.
Because this tendency is strong enough, ingroup favoritism occurs in the JG
group.
Their model and ours are consistent with each other because, when different
groups can adopt different assignment rules, both their model and ours with
sufficiently many groups predict bistability between ST and JG.
Their model and ours complement each other in the following respects.
First, they investigated competition between assignment rules, whereas we mainly
studied the case in which all the groups share an assignment rule.
Second, they assumed a well-mixed population, whereas we varied the frequency of
ingroup and outgroup interactions.
Third, they studied competition among at most five groups (i.e., five assignment
rules), whereas we assumed a general number of groups.
%
%


\section*{Conclusion}
\label{sec:conclusion}

To explore the possibility of spontaneous ingroup favoritism in indirect
reciprocity, we analyzed a social dilemma game in a population with group
structure.
We showed that the degree of ingroup bias depends on the reputation assignment
rule.
In particular, considerable ingroup favoritism occurs under the so-called JG
assignment rule, whereby observers assign bad reputations to players helping bad
players.
Ingroup favoritism has been considered to be an evolutionary outcome~\cite{%
  Yamagishi1998,Yamagishi1999,Kiyonari2000,Yamagishi2008,Mifune2010}.
The present work supports this general idea.
To measure the dependency of ingroup bias on the assignment rule in behavioral
experiments may be an interesting challenge.


\section*{Appendices}

\appendix

\section{Numerical methods in the case of the homogeneous assignment rule}
\label{appendix:numerical-homo}


We prepare a population of $N = 10^3$ DISC players divided into $M$ groups of
equal size.
We consider an $N \times M$ reputation matrix, denoted by ${\bm R} = (r_{i,
\ell})$, where $r_{i, \ell} \in \{{\rm G, B, U}\}$ represents the reputation of
player $i$ ($1 \le i \le N$) in the eyes of the observer in group $\ell$ ($1 \le
\ell \le M$).
U represents the unknown reputation.
We assume that all the entries of ${\bm R}$ are equal to U in the beginning of a
run.
In a one-shot donation game, we randomly select a player $i$ as donor.
Then, with probability $\theta/(N/M-1)$, we select a recipient $j (\neq i)$ that
is in the donor's group.
With probability $(1-\theta)/(N-N/M)$, we select a recipient $j$ that is in a
group different from the donor's group.
When determining the action, the donor refers to $r_{j,k}$, where $k$ is the
donor's group.
We assume that the donor cooperates when $r_{j,k} = {\rm U}$.
After the game, the observer in each group $\ell$ ($1 \le \ell \le M$) assigns a
new reputation to donor $i$ such that $r_{i,\ell} = {\rm G}$ with probability
$\Phi_{\rm G}(a, r_{j,\ell})$ and $r_{i,\ell} = {\rm B}$ with probability $1 -
\Phi_{\rm G}(a, r_{j,\ell})$, where $a \in \{{\rm C, D}\}$ is the donor's action
and $\Phi_{\rm G}(a, r_{j,\ell})$ under each assignment rule is defined in
Tab.~\ref{tbl:assessment}.
When $r_{j,\ell} = {\rm U}$, we assume that the observer uses IM; $\Phi_{\rm
G}({\rm U}, {\rm C}) = 1-\mu$ and $\Phi_{\rm G}({\rm U}, {\rm D}) = \mu$.
We set $\mu = 0.01$.


After repeating $T = 10^5$ rounds of the donation game, we calculate the
fraction of G players in group $k$ in the eyes of the observer in group $\ell$,
which is given by $\Phi^*_{k, \ell}({\rm G}) = \sum_{i=1; {\rm\; player\;} i
{\rm \;in\; group\;} k}^N \delta(r_{i, \ell}) / (N/M)$, where $\delta({\rm G}) =
1$ and $\delta({\rm B}) = \delta({\rm U}) = 0$.
The fractions of G players in the eyes of ingroup and outgroup observers are
given by $\Phi^*_{\rm in}({\rm G}) = \sum_{k=1}^M \Phi^*_{k, k}({\rm G}) / M$
and $\Phi^*_{\rm out}({\rm G}) = \sum_{k=1}^M \sum_{\ell=1, \ell \neq k}^M
\Phi^*_{k, \ell}({\rm G}) / [M(M-1)]$, respectively.
By substituting these quantities in Eqs.~(\ref{eq:cooperativeness}) and
(\ref{eq:ingroup-bias}), we obtain $\psi$ and $\rho$.
We average $\psi$ and $\rho$ over $10^2$ runs of the simulation.


\section{Numerical methods in the case of the heterogeneous assignment rule}
\label{appendix:numerical-hetero}


%
To analyze heterogeneous populations, we assume that observers in groups $1, 2,
\cdots, m$ adopt JG and those in groups $m+1, m+2, \cdots, M$, adopt ST.
By applying the procedure explained in Appendix~\ref{appendix:numerical-homo},
we obtain the fraction of G players in group $k$ in the eyes of the observer in
group $\ell$, i.e., $\Phi^*_{k, \ell}({\rm G})$.
The probability that a donor in group $k$ helps a recipient is given by
\begin{equation}
  \psi_k = \theta \Phi^*_{k, k}({\rm G}) + (1-\theta)
  \frac{1}{M-1}\sum_{\ell=1, \ell \neq k}^M \Phi^*_{\ell, k}({\rm G}).
\end{equation}
The probability that a recipient in group $k$ is helped by a donor is given by
\begin{equation}
  \phi_k = \theta \Phi^*_{k, k}({\rm G}) + (1-\theta)
  \frac{1}{M-1}\sum_{\ell=1, \ell \neq k}^M \Phi^*_{k, \ell}({\rm G}).
\end{equation}
The ingroup bias of the players in group $k$ is given by
\begin{equation}
  \rho_k = \Phi^*_{k, k}({\rm G}) -
  \frac{1}{M-1}\sum_{\ell=1, \ell \neq k}^M \Phi^*_{\ell, k}({\rm G}).
\end{equation}
The payoff to the players in group $k$ is given by
\begin{equation}
  \pi_k = -c \psi_k + b \phi_k.
\end{equation}
The cooperativeness, ingroup bias, and payoff to the players in groups employing
JG and ST are defined by $Q_{\rm JG} = \sum_{k=1}^m Q_k / m$ and $Q_{\rm ST} =
\sum_{k=m+1}^M Q_k / (M-m)$, respectively, where $Q$ represents either $\psi$,
$\rho$, or $\pi$.
We average these quantities over $10^2$ runs for each parameter set to generate
Fig.~\ref{fig:equilibria-STJG}.


\section*{Competing interests}

The authors have no competing interests to declare.

\section*{Authors contributions}

MN and NM designed the model.
MN derived the analytical and numerical results.
MN and NM wrote the paper.

\section*{Acknowledgements}

We thank Shoma Tanabe for careful reading of the manuscript.
MN acknowledges the support provided through Grants-in-Aid for Scientific
Research (No.~10J08999) from JSPS, Japan.
NM acknowledges the support provided through Grants-in-Aid for Scientific
Research (No.~23681033) and Innovative Areas ``Systems Molecular
Ethology''(No.~20115009) from MEXT, Japan.

{\ifthenelse{\boolean{publ}}{\footnotesize}{\small}
  \bibliographystyle{bmc_article}
  \bibliography{refs} }


\ifthenelse{\boolean{publ}}{\end{multicols}}{}

\clearpage

\section*{Figures}


\begin{figure}[h!]
  \centering
  \includegraphics[width=85mm]{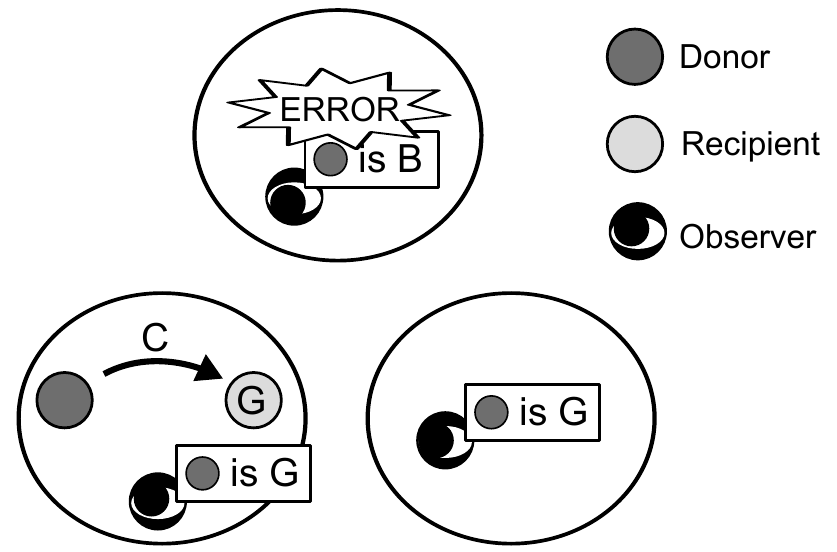}
  \caption{%
    {\bf Behavior of different observers in different groups ($M = 3$)}.
  }
  \label{fig:example-observation}
\end{figure}


\begin{figure}[h!]
  \centering
  \includegraphics[width=85mm]{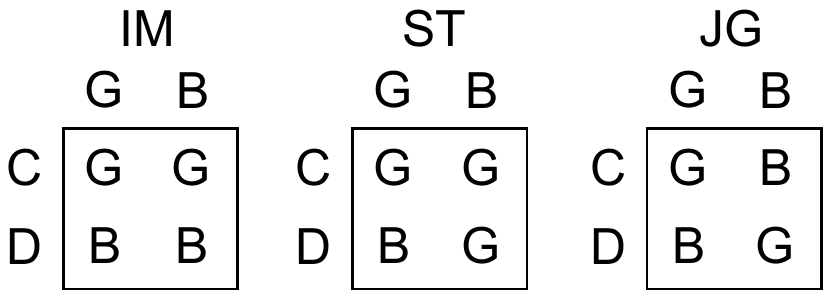}
  \caption{%
    {\bf Three reputation assignment rules}. Image scoring (IM),
    simple standing (ST), and stern judging (JG).
    The rows represent the donor's actions (i.e., C and D),
    the columns represent the recipient's reputations (G and B),
    and G and B inside the boxes represent the reputations that
    observers assign to the donor.
  }
  \label{fig:assessment}
\end{figure}


\begin{figure}[h!]
  \centering
  \includegraphics[width=85mm]{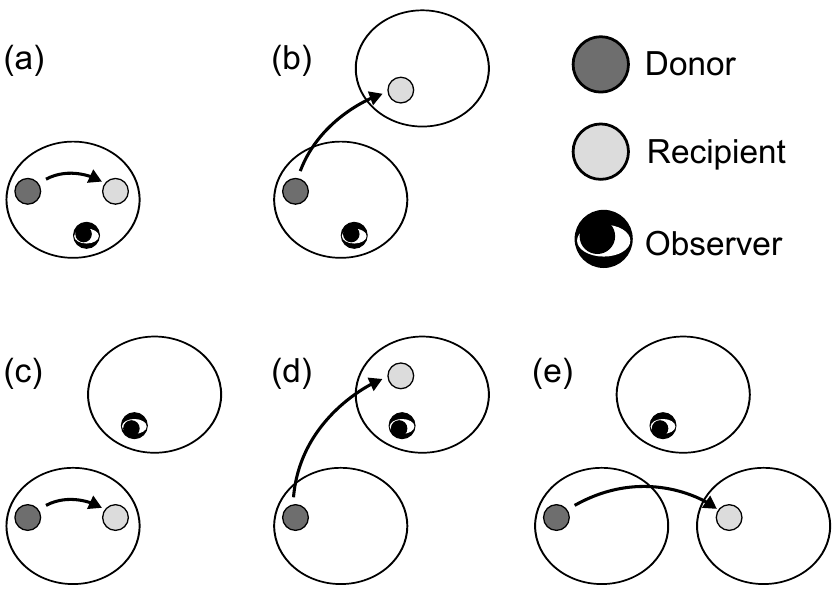}
  \caption{%
    {\bf Five possible situations of the reputation update}.
    %
    Observations are made by ingroup observers in {\bf (a)} and {\bf (b)},
    and by outgroup observers in {\bf (c)}, {\bf (d)}, and {\bf (e)}.
  }
  \label{fig:example-interactions}
\end{figure}


\begin{figure}[h!]
  \centering
  \includegraphics[width=85mm]{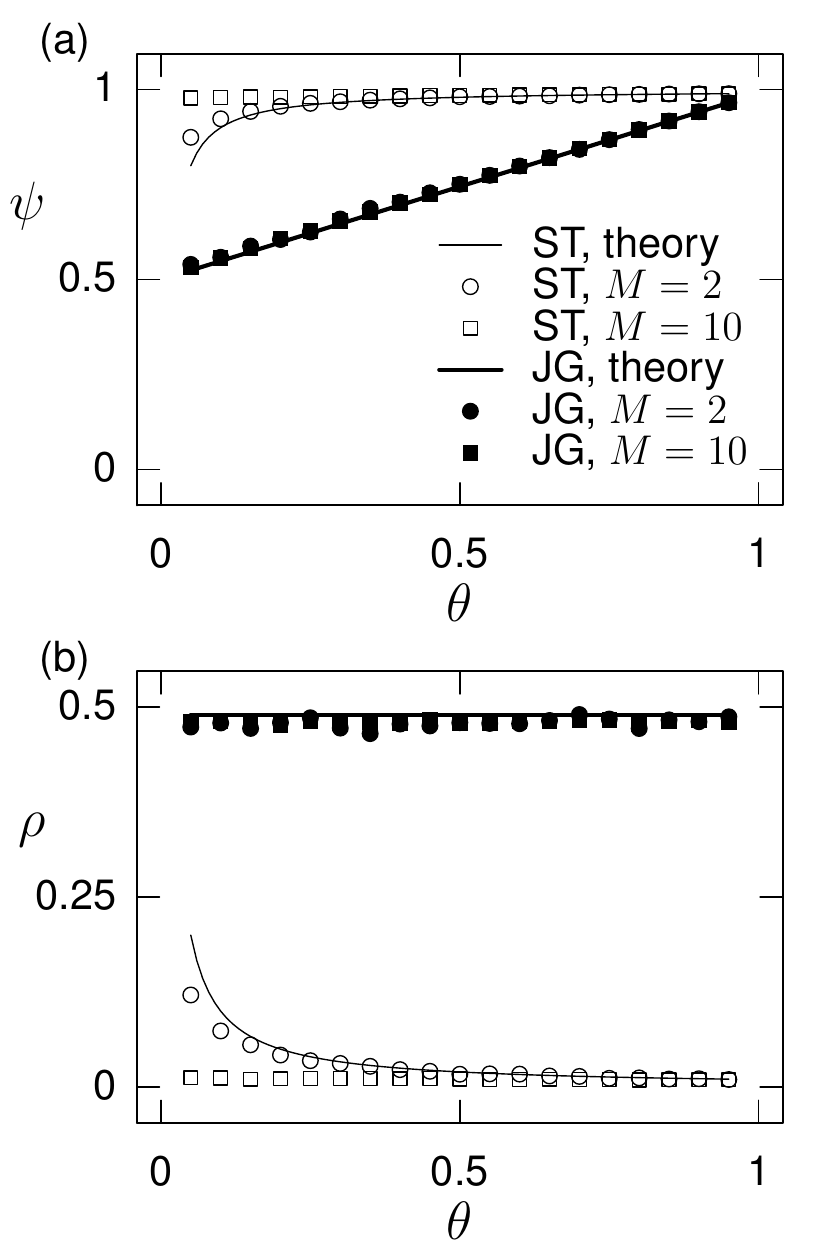}
  \caption{%
    {\bf Equilibria for a population of DISC players under ST and JG}.
    {\bf (a)} Cooperativeness ($\psi$) and {\bf (b)} ingroup bias ($\rho$).
    We vary the assignment rule (ST or JG),
    the number of groups ($M = 2$ or $10$), and
    the probability of ingroup interaction ($\theta$).
    The lines represent theoretical results shown
    in Tab.~\ref{tbl:equilibria}.
    The symbols represent numerical results.
  }
  \label{fig:equilibria}
\end{figure}


\begin{figure}[h!]
  \centering
  \includegraphics[width=85mm]{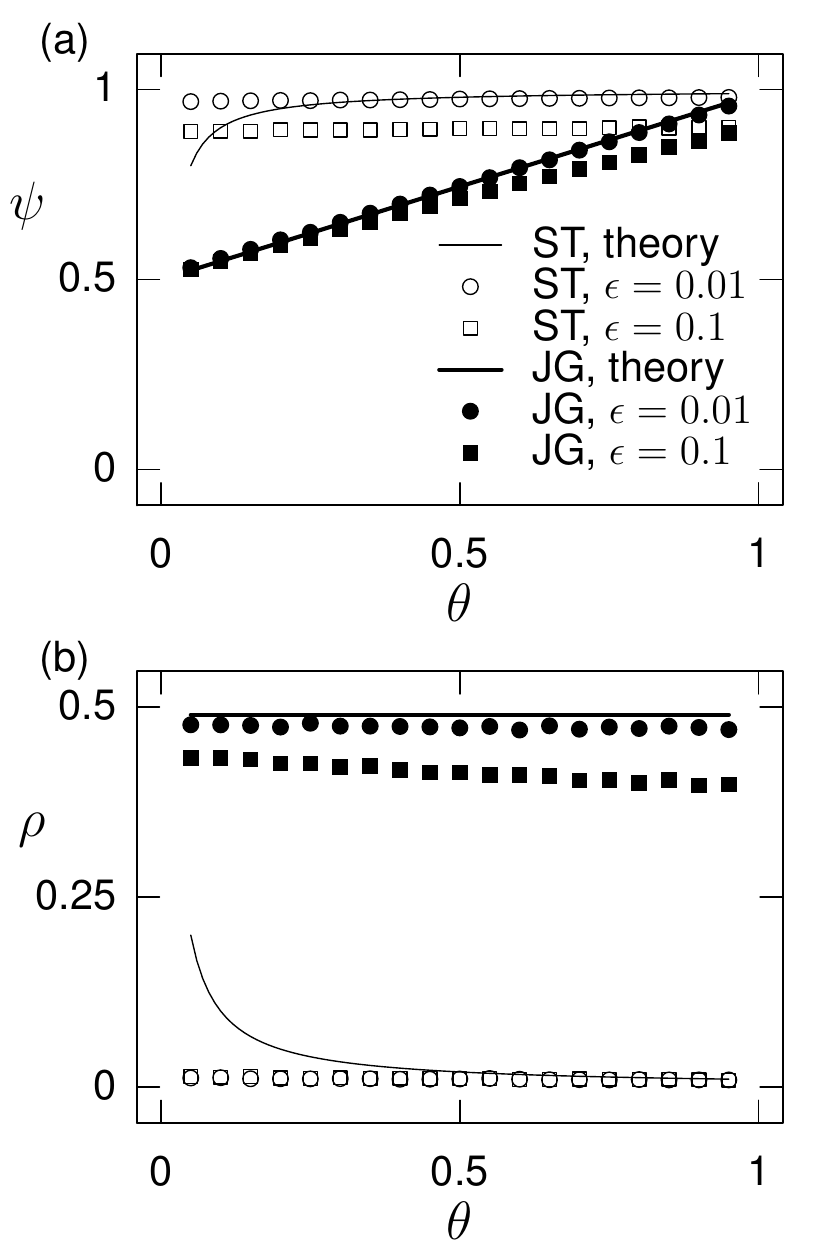}
  \caption{%
    {\bf Equilibria for a population of DISC players under action implementation
    error}.
    {\bf (a)} Cooperativeness ($\psi$) and {\bf (b)} ingroup bias ($\rho$).
    We fix the number of groups ($M = 10$) and vary the assignment rule (ST or
    JG), the probability that a donor fails to help a recipient ($\epsilon =
    0.01$ or $0.1$), and the probability of ingroup interaction ($\theta$).
    The lines represent theoretical results when $\epsilon = 0$ and are the
    replicates of those shown in Tab.~\ref{tbl:equilibria}.
    The symbols represent numerical results.
  }
  \label{fig:equilibria-eps}
\end{figure}


\begin{figure}[h!]
  \centering
  \includegraphics[width=170mm]{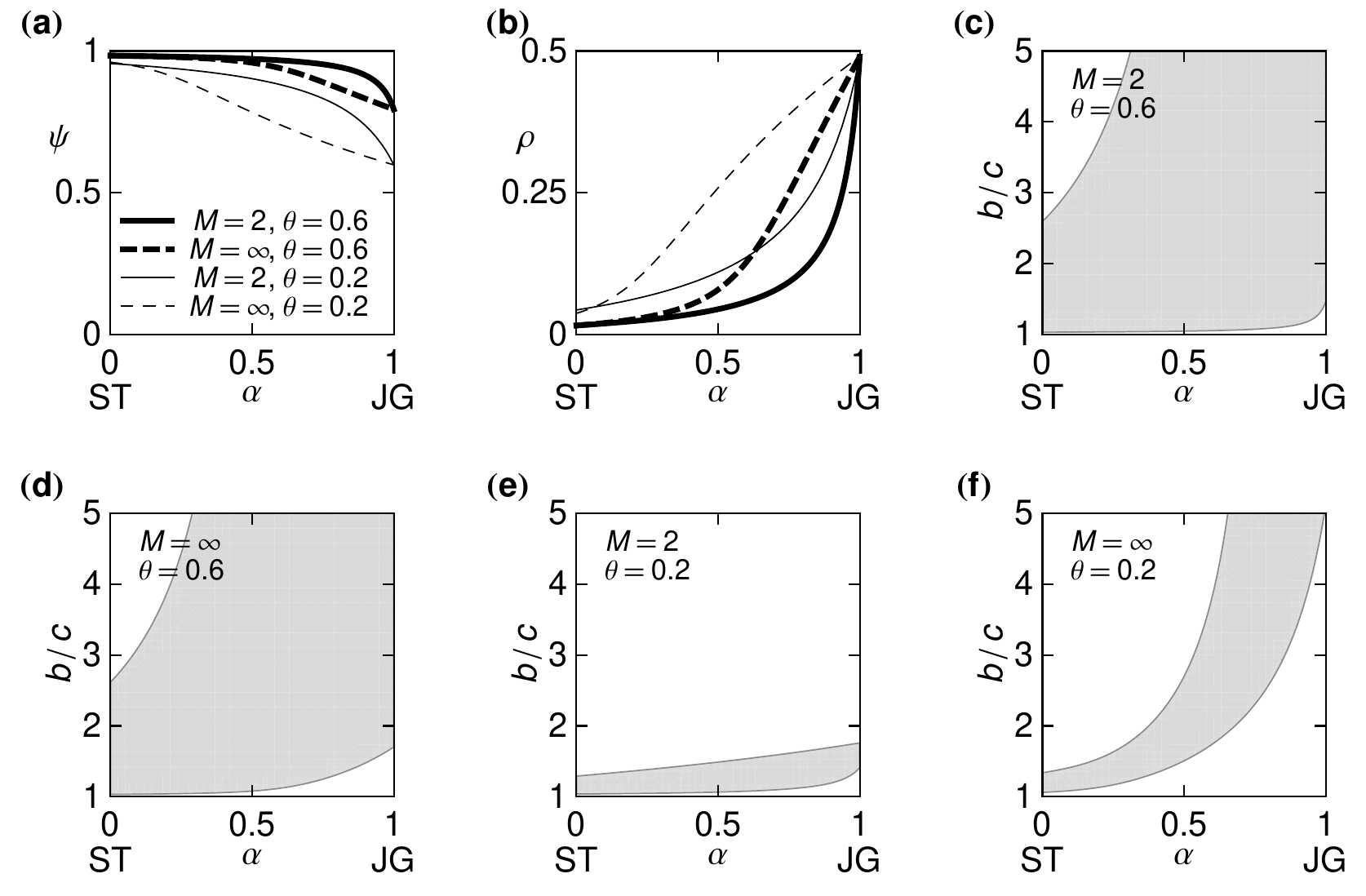}
  \caption{%
    {\bf Equilibria and the stability conditions for a population of DISC players
    under MX}.
    {\bf (a)} Cooperativeness ($\psi$) and {\bf (b)}  ingroup bias ($\rho$).
    In {\bf (a)} and {\bf (b)}, we set
    $(M, \theta) = (2, 0.6)$, $(\infty, 0.6)$, $(2, 0.2)$, and $(\infty, 0.2)$.
    {\bf (c)}\---{\bf (f)} Stability conditions.
    The homogeneous population of DISC players is stable in the shaded
    parameter regions.
    We set
    {\bf (c)} $(M, \theta) = (2, 0.6)$,
    {\bf (d)} $(M, \theta) = (\infty, 0.6)$,
    {\bf (e)} $(M, \theta) = (2, 0.2)$, and
    {\bf (f)} $(M, \theta) = (\infty, 0.2)$.
  }
  \label{fig:mixed}
\end{figure}


\begin{figure}[h!]
  \centering
  \includegraphics[width=170mm]{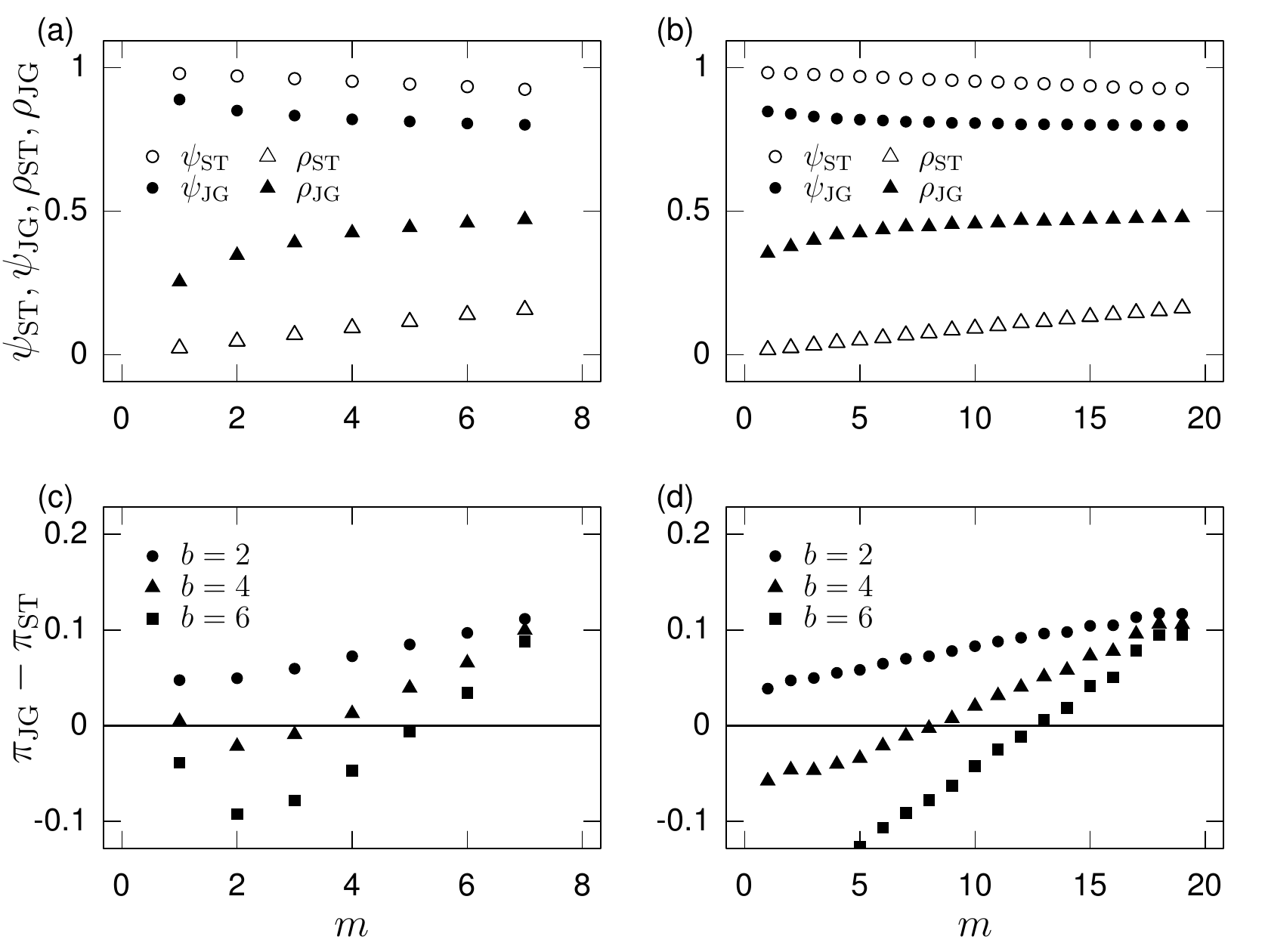}
  \caption{%
    {\bf Equilibria for a population of DISC players under heterogeneous
    assignment rules}.
    {\bf (a)}, {\bf (b)} Cooperativeness ($\psi_{\rm ST}$ and $\psi_{\rm JG}$)
    and ingroup bias ($\rho_{\rm ST}$ and $\rho_{\rm JG}$) for groups employing
    ST and JG.
    {\bf (c)}, {\bf (d)} Payoff difference between a player in a ST group and
    that in a JG group ($\pi_{\rm JG} - \pi_{\rm ST}$).
    We set $\theta = 0.6$ and $c = 1$.
    We also set $M = 8$  in {\bf (a)} and {\bf (c)},
                $M = 20$ in {\bf (b)} and {\bf (d)},
    and vary the number of JG groups (i.e., $m$) and $b$.
  }
  \label{fig:equilibria-STJG}
\end{figure}

\clearpage

\section*{Tables}


\begin{table}[h!]
  \centering
  \begin{tabular}{|l|p{37em}|}
    \hline
    Symbol & Meaning \\
    \hline
    $M$                            & Number of groups \\
    $\theta$                       & Probability that a donor and recipient in a one-shot game are in the same group \\
    ${\bm r} \in \{{\rm G, B}\}^M$ & Reputation vector of a player in the eyes of $M$ observers \\
    $p_k({\bm r})$                 & Probability that a player in group $k$ has reputation vector ${\bm r}$ \\
    $p_{-k}({\bm r})$              & Probability that a player outside group $k$ has reputation vector ${\bm r}$ \\
    $\sigma(r) \in \{{\rm C, D}\}$ & Donor's action to a recipient having reputation $r \in \{{\rm G, B}\}$ \\
    $\Phi_{r}(a, r^\prime)$        & Probability that an observer assigns reputation $r \in \{{\rm G, B}\}$ to a donor selecting action $a \in \{{\rm C, D}\}$ to a recipient having reputation $r^\prime \in \{{\rm C, D}\}$ \\
    $p_{\rm in}(r)$                & Probability that a player in the eyes of an ingroup observer has reputation $r$ \\
    $p_{\rm out}(r)$               & Probability that a player in the eyes of an outgroup observer has reputation $r$ \\
    \hline
  \end{tabular}
  \caption{%
    {\bf Meaning of symbols}.
  }
  \label{tbl:definitions}
\end{table}


\begin{table}[h!]
  \centering
  \begin{tabular}{|c|cccc|}
    \hline
    \makebox[2.6em][c]{Rule} &
    \makebox[5.2em][c]{$\Phi_{\rm G}({\rm C, G})$} &
    \makebox[5.2em][c]{$\Phi_{\rm G}({\rm D, G})$} &
    \makebox[5.2em][c]{$\Phi_{\rm G}({\rm C, B})$} &
    \makebox[5.2em][c]{$\Phi_{\rm G}({\rm D, B})$} \\
    \hline
    IM & $1-\mu$ & $\mu$ & $1-\mu$ & $\mu$ \\
    ST & $1-\mu$ & $\mu$ & $1-\mu$ & $1-\mu$ \\
    JG & $1-\mu$ & $\mu$ & $\mu$ & $1-\mu$ \\
    \hline
  \end{tabular}
  \caption{%
    {\bf Probability that an observer assigns G to a donor}.
    $\Phi_{\rm G}(a, r)$ represents the probability that a donor receives G
    when the donor selects action $a \in \{\rm C, D\}$ and the recipient has
    reputation $r \in \{\rm G, B\}$.
    The donor receives B with probability
    $\Phi_{\rm B}(a, r) = 1 - \Phi_{\rm G}(a, r)$.
  }
  \label{tbl:assessment}
\end{table}


\begin{table}[h!]
  \centering
  \begin{tabular}{|c|ccccc|}
    \hline
    \makebox[2.6em][c]{Rule} &
    \makebox[6.6em][c]{$\displaystyle p_{\rm in}^*({\rm G})$} &
    \makebox[6.6em][c]{$\displaystyle p_{\rm out}^*({\rm G})$} &
    \makebox[9.9em][c]{Stability condition} &
    \makebox[6.6em][c]{$\psi$} &
    \makebox[6.6em][c]{$\rho$} \\
    \hline
    & & & & & \\
    IM &
    $\displaystyle \frac{1}{2}$ &
    $\displaystyle \frac{1}{2}$ &
    Unstable &
    $\displaystyle \frac{1}{2}$ &
    $\displaystyle 0$ \\
    & & & & & \\
    ST &
    $\displaystyle 1-\mu$ &
    $\displaystyle 1 - \mu\frac{1+\theta}{\theta} + O(\mu^2)$ &
    Eq.~(\ref{eq:stability-condition-ST}) &
    $\displaystyle 1 - \frac{\mu}{\theta} + O(\mu^2)$ &
    $\displaystyle \frac{\mu}{\theta} + O(\mu^2)$ \\
    & & & & & \\
    JG &
    $\displaystyle 1-\mu$ &
    $\displaystyle \frac{1}{2}$ &
    Eq.~(\ref{eq:stability-condition-JG}) &
    $\displaystyle \frac{1+\theta}{2} - \mu\theta$ &
    $\displaystyle \frac{1}{2} - \mu$ \\
    & & & & & \\
    \hline
  \end{tabular}
  \caption{%
    {\bf Equilibria and the stability conditions for a population of DISC players
      under different assignment rules}.
  }
  \label{tbl:equilibria}
\end{table}

\end{bmcformat}
\end{document}